\documentclass[%
 reprint,
 amsmath,amssymb,
 aps,
]{revtex4-2}

\usepackage{graphicx}
\usepackage{dcolumn}
\usepackage{bm}
\usepackage{braket}

\begin{document}

\preprint{APS/123-QED}

\title{Sensing Aharonov--Bohm phase using a multiply-orbiting-ion interferometer}

\author{Ryoichi Saito$^{1}$}
\email{r-saito@phys.titech.ac.jp}
\author{Takashi Mukaiyama$^{1}$}%
\email{mukaiyama@phys.titech.ac.jp}
\affiliation{
$^1$Department of Physics, Tokyo Institute of Technology, Ookayama 2-12-1, Meguro-ku, 152-8550, Tokyo, Japan\\
}%

\date{\today}

\begin{abstract}
Interferometers, which are built using spatially propagating light or matter waves, are commonly used to measure physical quantities.
These measurements are made possible by exploiting the interference between waves traveling along different paths. This study introduces a novel approach to sensing of the Aharonov--Bohm phase, an ion matter-wave interferometer operating within a two-dimensional circular trajectory in a trap potential.
The ion orbitals in the potential form Lissajous curves, causing the direction of ion rotation to reverse. This reversal results in a corresponding change in the interference phase. 
Our study is groundbreaking as it is the first attempt to utilize propagating matter waves of an ion in constructing an interferometer for the measurement of physical quantities.
Given that  the scale factor of the interferometer to the cyclotron motion and the rotation of the system is common, 
the sensitivity to the Aharonov--Bohm phase in this study corresponds to a rotation sensitivity of approximately 300~rad/s.
Besides advancing interferometry, our work also lays the foundation for future research into the use of ion matter waves in gyroscopic applications.
\end{abstract}

\maketitle


Quantum sensing, a practice of detecting physical quantities by leveraging their quantum mechanical properties, has flourished in recent years. This progress is largely attributed to technological improvements that allow for the precise manipulation of quantum systems~\cite{RevModPhys.89.035002}.
Among the many applications of quantum technology, sensing stands out as one of the most promising applications of quantum in society and is anticipated to reach practical implementation in the near future.

The broad array of physical quantities that can be measured using quantum sensing depends on the specific quantum system in question. These systems interact with different physical quantities, which they measure. Examples of highly reproducible quantum systems include flying atoms, cooled and trapped atomic ensembles, and ions. To date, numerous methods for manipulating these states have been developed, making them ideal platforms for quantum sensing. For instance, cooled atomic systems are excellent for sensing magnetic fields~\cite{PhysRevLett.98.200801, PhysRevLett.111.143001, PhysRevLett.109.253605}, electric fields~\cite{Zeiske1995, Phys.Rev.A51R4305, PhysRevA.65.042104}, molecular properties~\cite{PhysRevA.51.3883, PhysRevLett.74.1043}, time~\cite{Takamoto2005, doi:10.1126/science.1240420, Bloom2014} and inertial forces~\cite{PhysRevLett.67.181, PhysRevA.54.3165, Nature4008491999, PhysRevLett.78.2046, ClassicalQuantumGravity17, PhysRevApplied.12.014019}.
Trapped ions have also proven to be effective platforms for detecting forces~\cite{Biercuk2010, Ivanov2016, doi:10.1126/sciadv.aao4453, Shaniv2017, 10.1063/5.0046121, PhysRevApplied.16.044007}, time~\cite{PhysRevLett.116.063001, 10.1126/science.1114375}, and electric and magnetic fields~\cite{Kevin2021, PhysRevA.104.052610, PhysRevX.7.031050, Noguchi2014}.

Atomic interferometers~\cite{RevModPhys.81.1051} designed to detect inertial forces typically use a Mach-Zehnder-type interference configuration ~\cite{PhysRevLett.67.181, PhysRevA.54.3165, Nature4008491999, PhysRevLett.78.2046, ClassicalQuantumGravity17, PhysRevApplied.12.014019}. This set-up spatially divides atomic matter waves, allowing for their temporal evolution before they are recombined. During this process, external perturbations, such as acceleration and rotation, cause distinct phase shifts along different matter wave paths. The magnitude of these external perturbations can be gauged by measuring changes in the final quantum state population. This change arises from the phase difference between the two quantum states in the quantum superposition throughout their temporal evolution. Therefore, by evaluating the change in the final quantum state population, we can accurately measure the magnitude of external perturbations in atomic interferometers.

Despite the advancements in quantum sensing, there are no existing instances of using a spatially propagating matter-wave interferometer of a trapped ion for this purpose.
This delay is attributed to technical challenges, primarily the difficulty of creating large spatial interference areas using strongly confined ions in a trap. Consequently, the development of spatially propagating matter-wave interferometry in ion trap systems has trailed behind that of atomic systems.
However, ion trap systems can sustain stable interferometers for long periods due to their exceptionally deep trapping potentials.
In contrast to beam-based interferometers, trap systems can generate interferometers featuring multiple orbits.
The operating principle of this interferometry may facilitate long-term integration of the interference area across multiple orbits, resulting in a large interference area while maintaining a compact device volume.
Focusing on these advantages, a recent proposal suggested a protocol for rotation sensing that uses a spatially propagating ion matter wave in a trap potential ~\cite{Campbell_2017}.
Remarkably,  ion gyroscopes are insensitive to acceleration since the typical translational acceleration meets the adiabatic condition for a simple trap-center drift. This drift does not induce a phase shift in the interferometer.
This represents an advantage absent in inertial sensors employing atomic beams.

\begin{figure*}[t]
	\centering
	\includegraphics[width=17cm]{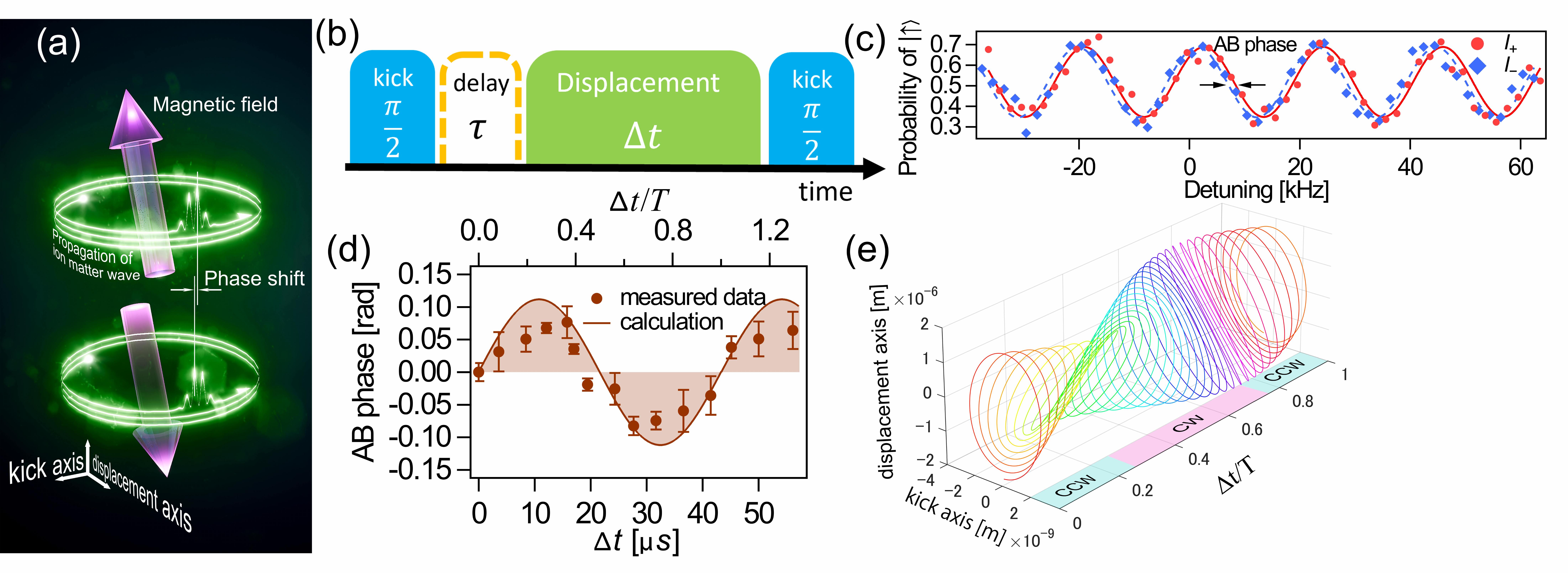}
	\caption{
	(a): Schematic of an ion matter-wave interferometer for AB phase detection.
	(b): Experimental procedure for measuring the AB phase.
	(c): Typical interference signal acquired by tuning the detuning for the Raman transition. The AB phase is determined by analyzing the signal obtained with forward and backward current flowing through the external coil.
	(d): Time-dependent AB phase of the ion matter–wave interference signal.
	(e): Time-dependent two-dimensional orbit of the ion matter-wave at $\tau = 0$.
	The ion orbit rotates counterclockwise in the first quarter period, rotates clockwise in the subsequent half period, and reverts to a counterclockwise motion in the final quarter period.
	}
	\label{fig:fig1}
\end{figure*}

In this study, we have successfully demonstrated matter-wave interferometry of a trapped ion and detected the Aharonov--Bohm (AB) phase.
We achieved this by generating an ion matter-wave interferometer, created from a wave packet of an ion in circular motion within a nearly isotropic two-dimensional trapping potential.
Our results confirm the linear dependence of the phase shift in the interference fringe on both the magnetic field strength and the effective interference area. The latter is determined by the multiple circulations of the ion motion.
This research is groundbreaking as it represents the first measurement of physical quantities using spatial ion matter-wave interferometry.
In fact, the sensitivity of the interferometer to the cyclotron motion and the rotation of the system are common~\cite{Campbell_2017, footnote}.
The sensitivity to the AB phase realized in this study corresponds to the rotation sensitivity of approximately 300~rad/s.
Our study advances the field of quantum sensing and lays the groundwork for the practical implementation of matter-wave interferometry of ions.

To detect the AB phase, we use a matter-wave interferometer to generate a nearly circular interference area. An ion is put in a superposition of clockwise and counterclockwise rotation states to construct the interferometer. The AB shift $\phi$ is described by the following equation~\cite{Campbell_2017},
\begin{equation}
\label{eq:eq1}
\phi = \frac{e}{\hbar}\int_C \vec{A}\cdot {\rm d}\vec{l} = \frac{e \omega}{2 \pi} B_{\perp} \int_0^{\Delta t}S(t) {\rm d}t.
\end{equation}
In this equation $e, \omega/2\pi, \hbar, B_{\perp}, \Delta t$ represent the elementary charge, trap frequency (also considered as the ion’s orbiting frequency), Dirac constant, external magnetic field normal to the ion orbit plane, and interference time, respectively. $S(t){\rm d}t$ represents the area that the vector from the trap center to the ion position sweeps in the time duration of ${\rm d}t$.

Figure \ref{fig:fig1}(a) depicts a conceptual sketch of the experimental set-up. We induce an ion’s rotational orbit through a sequential process that involves a momentum kick driven by a Raman transition using a mode-locked pulsed laser, and a trap center displacement applied in an orthogonal direction to the momentum kick axis. We achieve this displacement by applying a step voltage to one of the ion trap electrodes\cite{Alonso2016, PhysRevA.104.053114, 10.1063/5.0100007}. The AB phase is quantified by subtracting the phase values measured by applying opposite magnetic field directions. This approach enables us to detect differences in the interference phase solely attributed to variations in the applied magnetic field direction.

A more detailed experimental procedure is outlined in Fig.\ref{fig:fig1}(b).
We start by trapping a single $^{171}\rm Yb^{+}$ ion in a conventional linear Paul trap and apply standard Doppler cooling. The trap frequencies are fine-tuned to create a nearly two-dimensional isotropic potential, typically 800 kHz in two directions, with a biaxial difference frequency of $20$ to $25~\rm kHz$. The residual trap frequency is approximately 1.7 times higher (see Methods), resulting in the ion being confined in a pancake-like potential. We use the internal states $\ket{F=0, m_F=0} = \ket{\downarrow}$ and $\ket{F=1, m_F=0} = \ket{\uparrow}$ in the hyperfine ground states of the $^{171}\rm Yb^{+}$ ion as qubit states. Initially, the ion is prepared in the $\Ket{\downarrow}$ state, and then we apply a half $\pi$ pulse of Raman transition between $\Ket{\downarrow}$ and $\Ket{\uparrow}$ along one of the weakly confining directions of the pancake potential using a mode-locked pulsed laser. This generates an entangled state between the internal and momentum states\cite{PhysRevLett.105.090502, PhysRevLett.104.140501, PhysRevLett.110.203001}, constructing a one-dimensional matter-wave interferometer\cite{PhysRevLett.126.153604}. Subsequently, the trap center displacement is applied along the orthogonal direction of the momentum kick axis after a time delay $\tau$\cite{Alonso2016, PhysRevA.104.053114, 10.1063/5.0100007}.
The ion in the $\Ket{\uparrow}$ state, kicked by the photons, moves in an elliptical motion, thereby creating an effective area for sensing the AB phase in the trap potential. By contrast, the ion in the $\Ket{\downarrow}$ state moves linearly along the direction of the trap-center displacement without creating an effective interference area for detecting the AB phase.
After the trap center displacement with the time duration of $\Delta t$, we return the trap center to its original position. Finally, we apply another half $\pi$ pulse to the ion and measure the interference fringes of single-ion matter-wave interferometry. Following this procedure, the population in the $\Ket{\uparrow}$ qubit states is read using standard fluorescence detection.

Figure~\ref{fig:fig1}(c) depicts a typical two-dimensional matter-wave interference signal obtained from the aforementioned procedure. The horizontal axis represents the detuning from the resonance of the Raman transition, which is controlled by adjusting the frequency difference between the counterpropagating beams. The vertical axis indicates the probability of detecting the ion in the upper state after the interference protocol. The interference fringe with a forward current $I_+$ applied to the external bias coil is plotted with red circles, while the interference fringes with a backward current $I_-$, obtained by flipping the direction of the current in the coil, are marked with blue diamonds. For this measurement, we set the displacement time $\Delta t$ to $20.9~\rm \mu s$, the applied magnetic field to $6.96~\rm Gauss$, and the delay time to $\tau = 3T_{\text{kick}}/4$, with $T_{\text{kick}}$ representing the trapping period of the ion along the momentum kick axis. The solid and dashed sinusoidal curves depicted in the figure represent the results of fitting to the interference fringes of the forward current $I_+$ and backward current $I_-$, respectively. The interference fringes shown in Fig. \ref{fig:fig1}(c) correspond to the Ramsey interference fringes, whose period aligns with the reciprocal of the time between half $\pi$ pulses. The AB phase is measured as the phase difference caused by magnetic field reversal. 
The magnetic field strengths in the forward and backward currents in the coil are calibrated by microwave spectroscopy (see Methods). 
The phase shift caused by the slight difference in the magnetic field strength is eliminated by subtracting the phase shift values measured without the trap center displacement.

We measured the AB phase with varying displacement times $\Delta t$, corresponding to the two-dimensional interference time, as shown in Fig.~\ref{fig:fig1}(d). The vertical axis illustrates the measured AB phase, while the horizontal displacement duration. The difference in the magnetic field induced by reversing the current applied to the external bias coil is $2 B_{\perp}=13.9~\rm Gauss$. Here, we set the delay time $\tau = 0$. Each point in Fig.~\ref{fig:fig1}(d) is obtained from an average of typically seven AB phase measurements, as depicted in Fig.~\ref{fig:fig1}(c). Error bars represent the standard error. The AB phase exhibits an increase or decrease rather than a monotonic change depending on the interference time $\Delta t$.Fig.~\ref{fig:fig1}(e) illustrates the ion trajectories generated by using the aforementioned method, with a delay time $\tau = 0$.
The kick and displacement axes depict the directions in which the momentum kick and trap center displacement are applied. The remaining axis, denoted as $T$, represents the time axis normalized by the reciprocal trap frequency difference between the kick and displacement axes. Owing to the trap not being perfectly isotropic, the ion’s rotational orbit forms a Lissajous curve rather than a closed elliptic orbit, leading to a temporal change in its interference area. As depicted in Fig.~\ref{fig:fig1}(e), the period of the Lissajous ion orbit aligns with $T$, the reciprocal of the biaxial difference frequency. Notably, the ion orbit rotates counterclockwise in the initial $T/4$ period. In the succeeding half-period, the rotation direction reverses to clockwise, and in the last quarter period, it reverts back to counterclockwise, as indicated along the time axis of Fig.~\ref{fig:fig1}(e). This reversal in the rotation direction also inverses the change of the AB phase. We calculate the measured AB phase over time, as depicted by the solid curve in Fig.~\ref{fig:fig1}(d), using Eq.~\ref{eq:eq1}. The top axis of Fig.~\ref{fig:fig1}(d) represents the time normalized by $T$. Notably, the experimental results align closely with the theoretical curve.

\begin{figure}[t]
	\centering
	\includegraphics[width=5.5cm]{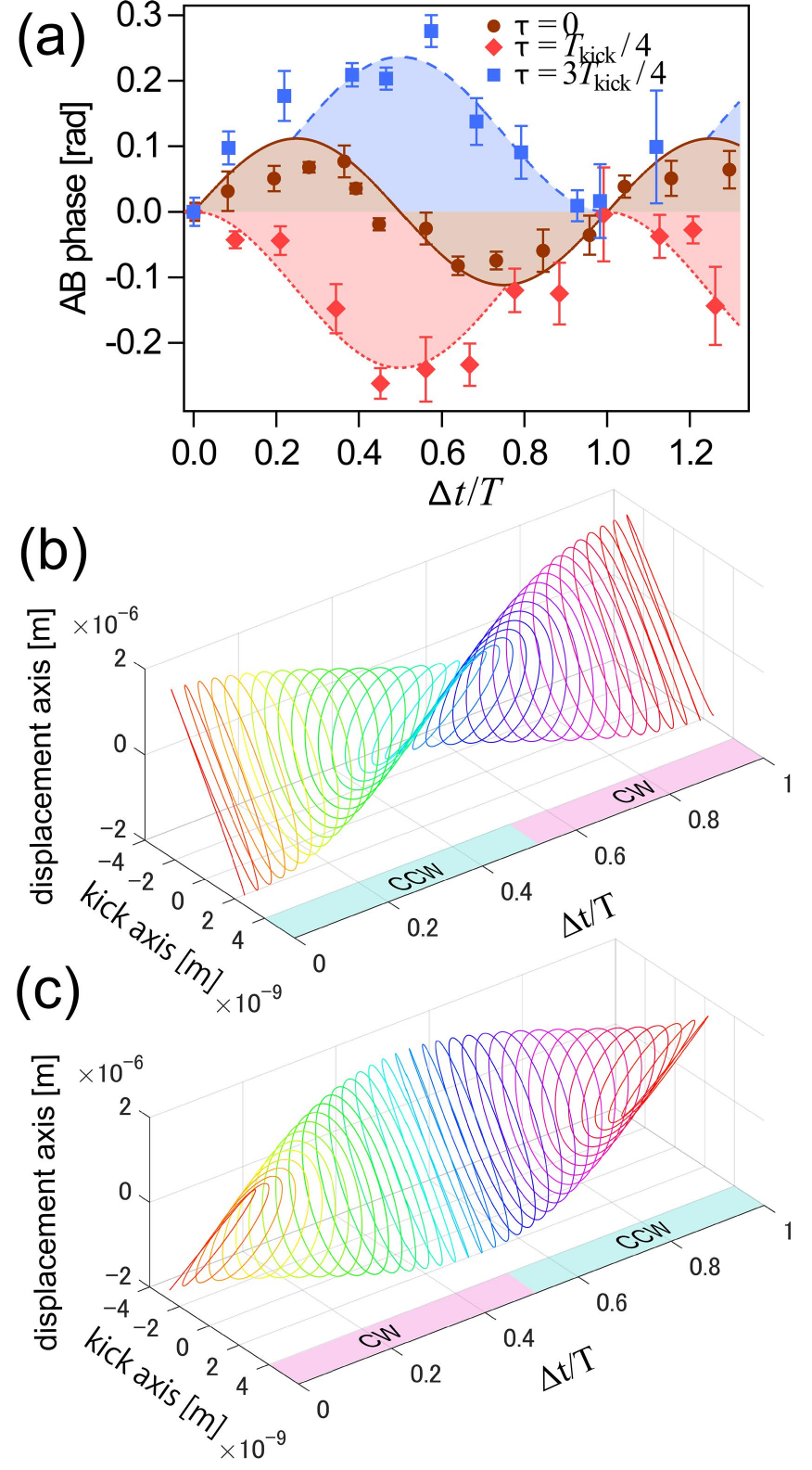}
	\caption{
	(a): The time-dependent AB phase at three different delay times $\tau$.
	In panels (b) and (c), the calculated rotational orbits of the ion in the nearly two-dimensional isotropic potential are shown for $\tau = T_{\rm kick}/4 $ and $3T_{\rm kick}/4$, respectively.
	}
	\label{fig:fig2}
\end{figure}

We have investigated the time-dependence behavior of the measured AB phase by varying the delay times $\tau$, thereby modifying the initial condition of the ion orbit in the trap potential. Figure \ref{fig:fig2}(a) presents this time dependence under three distinct delay times. The brown circles correspond to $\tau = 0$ and are identical to those depicted in Fig.\ref{fig:fig1}(d). The blue squares represent $\tau = T_{\rm kick}/4$, whereas the red diamonds denote $\tau = 3T_{\rm kick}/4$. The solid, dashed, and dotted curves illustrate the AB phase calculated according to Eq.~\ref{eq:eq1}. This dependence on delay time can be comprehended by examining the ion rotational orbits corresponding to different delay times. Figures~\ref{fig:fig2}(b) and c show the calculated ion rotational orbits within the nearly isotropic two-dimensional potential for $\tau = T_{\rm kick}/4$ and $3T_{\rm kick}/4$, respectively.
For $\tau = T_{\text{kick}}/4$, the rotation direction is counterclockwise during the first half period and clockwise during the latter half period, as depicted in Fig.~\ref{fig:fig2}(b),.
Similarly, the AB phase increases during the first half period, reaching its maximum at the midpoint, and then decreases during the second half period, as indicated by the blue dashed curve in Fig.~\ref{fig:fig2}(a).
By contrast, for $\tau = 3T_{\rm kick}/4$, the ion revolves clockwise during the first half period and counterclockwise during the following half period, as illustrated in Fig. \ref{fig:fig2}(c). Here, the AB phase decreases during the first half period, reaches a minimum at the half period, and then increases during the latter half period. Notably, for both $\tau = T_{\rm kick}/4$ and $3T_{\rm kick}/4$, the absolute value of the measured AB phase is maximized, thereby enhancing sensitivity to the AB phase

\begin{figure}[tbp]
	\centering
	\includegraphics[width=8.5cm]{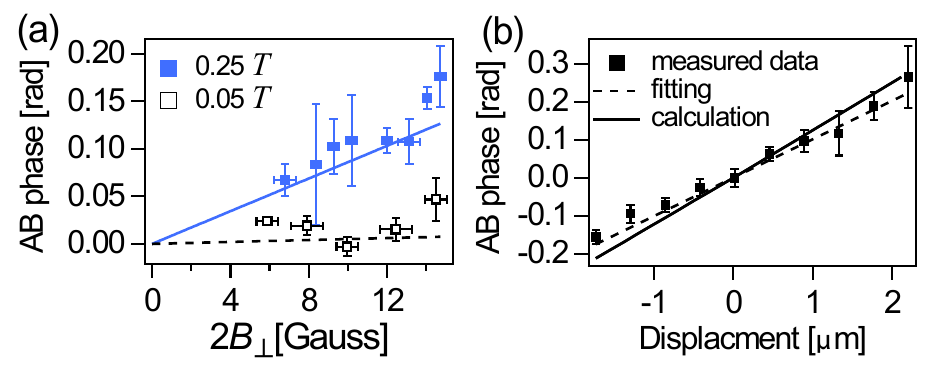}
	\caption{
	(a): Magnetic field dependence of the AB phase of ion matter wave interference at delay time $\tau = T_{\rm kick}/4 $.
	(b): Displacement size dependence of the AB phase measured with delay time $\tau = T_{\rm}$ and displacement time $\Delta t = T/2$.
	}
	\label{fig:fig3}
\end{figure}

We then investigate the magnetic field dependence of the AB phase while keeping the delay time $\tau$ and displacement time $\Delta t$ constant. Figure~\ref{fig:fig3}(a) demonstrates the magnetic field dependence of the AB phase as a function of the applied magnetic field for a delay time of $T_{\rm kick}/4$. The horizontal axis represents the magnetic field difference $2B_{\perp}$. The blue-filled and black-opened squares illustrate the measured AB phase for displacement times of $\Delta t = 0.25T$ and $0.05T$, respectively. The results indicate that for $\Delta t = 0.25T$, a consistent linear dependence exists on the magnetic field. Conversely, for $\Delta t = 0.05T$, the AB phase demonstrates minimal dependence on the magnetic field owing to the reduced interference area. The blue solid and black dashed lines depict the calculated AB phase for varying magnetic fields, aligning closely with the experimental data.

Finally, we measured the AB phase by varying the displacement size while maintaining consistent delay time, displacement time, and magnetic field.
Figure~\ref{fig:fig3}(b) presents the experimental result for $\tau =T_{\rm kick}/4$, $\Delta t = T/2$, and $B_{\perp} = 13.9~\rm Gauss$. As observed in Fig.~\ref{fig:fig2}(a), the conditions of $\tau =T_{\rm kick}/4$ and $\Delta t = T/2$ yield the maximum sensitivity to the AB phase.
The vertical axis in Fig.~\ref{fig:fig3}(b) depicts the measured AB phase, whereas the horizontal axis illustrates the displacement size controlled by the step voltage applied to the ion trap electrode. 
The integrated interference area in Eq.~\ref{eq:eq1} is proportional to the displacement size. Consequently, the AB phase is expected to exhibit a linear dependency on the displacement size.
The solid and dashed lines in Fig.~\ref{fig:fig3}(b) represent the theoretically expected AB phase and the fitting result of the data points to a linear function, respectively.
The results from the negative displacement region in Fig.~\ref{fig:fig3}(b) are obtained by reversing the direction of the trap center displacement using an opposite step voltage. This reversal leads to a change in the direction of the ion rotation, causing the measured AB phase to switch its sign, as evident in Fig.~\ref{fig:fig3}(b).

In conclusion, we successfully implemented a matter-wave interferometer using a trapped ion in a harmonic potential. The AB phase is detected using an ion matter wave in a superposition of circularly rotating and linearly moving states. 
The sensitivity to the AB phase realized in this study corresponds to the rotation sensitivity of approximately 300~rad/s~\cite{Campbell_2017, footnote}.
This marks a significant advancement toward realizing a gyroscope with a trapped ion. 
This study represents the first-ever attempt to use a two-dimensional interferometer constructed by an ion wave packet propagating in a circular potential. This achievement holds promise for future development in quantum sensing, leveraging the inherent quantum nature of wave phenomena.
However, our current interferometer’s sensitivity is limited owing to the integration time. In our experimental set-up, the trap potential is not perfectly circular, causing the ion to follow a Lissajous curve in which the rotation direction reverses, leading to the cancelation of the interference signal over time. To enhance the interferometer sensitivity, it is crucial to achieve equal trap frequencies in both axes. Realizing an interference time of 1~ms necessitates suppressing the trap frequency difference between the two axes to 0.5~kHz, a considerable reduction compared to the smallest trap frequency difference of $5~\rm kHz$ observed in our trap set-up. Misalignment of ion trap electrodes induces coupling between the two orthogonal trap axes, resulting in a nonzero trap frequency difference. A meticulously designed ion trap structure is essential to mitigate this issue and extend the interrogation time for the matter-wave interferometer. Furthermore, enhancing the interference area by increasing the displacement size and employing multiple momentum kicks~\cite{Johnson2017}, can further improve performance.

We acknowledge fruitful discussion with Mikio Kozuma.
This work is supported by JST-Mirai Program Grant Number JPMJMI17A3, Co-creation place formation support program (JPMJPF2015) and JSPS KAKENHI Grant Numbers JP23K13046.



\begin{thebibliography}{9}
\bibitem{footnote} The observed phase is described as $ \Phi = \Phi_{\rm rot}+ \Phi_{\rm cyc} = \frac{4 m \omega A_0}{h} \Delta t \left( \Omega_{\rm  rot}/2+\Omega_{\rm cyc} \right)$. Here,  $A_0$ is an interference area, $\Omega_{\rm rot}$  and $\Omega_{\rm cyc}$ are the angular frequencies of system rotation and cyclotron oscillation, respectively. The scale factor $\frac{\partial \Phi}{\partial \Omega}$ is common for a cyclotron motion and system rotation.
\end{thebibliography}

\end{document}